\documentclass[preprint,aps]{revtex4}

\newcommand{\be}{\begin{equation}} \newcommand{\ee}{\end{equation}} 
\newcommand{\bea}{\begin{eqnarray}}\newcommand{\eea}{\end{eqnarray}}

\begin{document}
\preprint{SINP/TNP/04-04, hep-th/0403083}
\title{ Supersymmetric quantum mechanics on noncommutative space}
\author{ Pijush K. Ghosh}
\email{pijush@theory.saha.ernet.in}
\affiliation{Theory Division, Saha Institute of Nuclear Physics,\\ 
Kolkata 700 064, India.}
\begin{abstract} 
We construct supersymmetric quantum mechanics in terms of two real
supercharges on noncommutative space in arbitrary dimensions. We obtain
the exact eigenspectra of the two and three dimensional noncommutative
superoscillators. We further show that a reduction in the phase-space occurs
for a critical surface in the space of parameters. At this critical surface,
the energy-spectrum of the bosonic sector is infinitely degenerate, while the
degeneracy in the spectrum of the fermionic sector gets enhanced by a factor
of two for each pair of reduced canonical coordinates. For the two dimensional
noncommutative `inverted superoscillator', we find exact eigenspectra with a
well-defined groundstate for certain regions in the parameter space, which have
no smooth limit to the ordinary commutative space. 
\end{abstract}
\maketitle
\newpage
\section{introduction}

The noncommutative space or the associated algebra arises in many branches
of physics and mathematics\cite{connes}. The reason behind the recent upsurge
in the study of field theory on the noncommutative space is that such theories
naturally appear in a particular low-energy limit of open string theory on
D-brane backgrounds in the presence of a constant antisymmetric tensor
field\cite{sw}. There are already many interesting results on different aspects
of both perturbative\cite{pert} and non-perturbative\cite{npert}
non-commutative field theory\cite{rev1,rev2,rev3}. The study of supersymmetric
field theories on noncommutative spaces with the standard
anticommuting\cite{bra} or non-anticommuting\cite{tama,seib} odd coordinates
is also an active area of research.

Along the same line of development, quantum mechanics on the noncommutative 
space has been studied
extensively\cite{fr,lamb,masud,poly,3d1,others,rabi,3d2,3d3,invert}. The
complete eigenstates of the noncommutative oscillator\cite{poly,3d1,3d2,3d3}
have been found analytically in two and higher dimensions. In particular, the
spectrum of the noncommutative oscillator is shown to be identical to that of
an anisotropic oscillator on the corresponding commutative space. The
frequencies of the anisotropic oscillator are, in general, a function of the
parameters appearing in the noncommutative algebra of the phase-space
operators. In the appropriate commutative limit, the spectrum of an isotropic
oscillator on commutative space is recovered. No other nontrivial quantum
mechanical problem is known to be completely solvable on the noncommutative
plane or higher dimensions.

Supersymmetric quantum mechanics on commutative space is a well developed
subject\cite{cks}. The study of supersymmetric quantum mechanics enriched us in 
understanding many subtle issues of supersymmetric field theory in a much
simpler way. Further, different ideas and techniques emerging from this
subject can be used to solve algebraically a host of quantum mechanical problems
arising in diverse branches of physics, including those found in any
standard textbook on quantum mechanics. The $\star$-product formulation of
supersymmetric quantum mechanics on commutative space is also known\cite{star}.

It is natural to enquire at this juncture about the present status of
the supersymmetric quantum mechanics on the noncommutative space. Recently,
the Pauli equation on the noncommutative plane was shown to be supersymmetric
with the gyro-magnetic ratio $g=2$\cite{khare}. The superoscillator on the
noncommutative plane has also been constructed\cite{pp}. However, we find that
a general and uniform formulation of the ${\cal{N}}=2$ supersymmetric quantum
mechanics on noncommutative space is lacking. For example, the supersymmetric
constructions in Refs. \cite{khare,pp} are not suitable, (i) for dimensions
$N \geq 3$, (ii) for describing quantum mechanical Hamiltonian with any
arbitrary superpotential and (iii) for accommodating more general
noncommutativities among phase-space operators.

The purpose of this paper is to present a formulation of supersymmetric quantum
mechanics on noncommutative space without the shortcomings discussed above.
In particular, we construct supersymmetric quantum mechanics in terms of
${\cal{N}}=2$ real supercharges on noncommutative space in arbitrary dimensions.
This construction is valid for any arbitrary superpotential and accommodates
a very large class of noncommutativities among phase-space operators.
We study the superoscillator on two and three dimensional noncommutative
spaces in some detail.

We obtain the complete spectrum of the two dimensional
superoscillator analytically. The spectrum is identical to that of an
anisotropic superoscillator on the commutative plane. 
The frequencies of the anisotropic superoscillators are functions of the
parameters appearing in the non-commutative algebra of the phase-space
operators. In the commutative limit, the isotropic superoscillator is
recovered. Further, we determine the spectrum of the `inverted superoscillator'
on the noncommutative plane with a well defined groundstate for certain
regions in the parameter space. There is no smooth limit to the commutative
space from these regions of the parameter space. Although, the Hamiltonian
of the `inverted superoscillator' on the noncommutative plane is non-hermitian
and not semi-positive definite, the spectrum obtained is indeed real and
positive.

A general
analysis of the three dimensional noncommutative superoscillator is messy
and complicated. So, we present the energy spectrum for a few specific
choices of the parameters. For all these cases, the energy spectrum is
identical to that of an anisotropic superoscillator on three dimensional
commutative space. The commutative limit for all these cases is smooth
and reproduces the well known results for usual isotropic superoscillator in
three dimensions.

We show that for both two and three dimensional noncommutative
superoscillators, a reduction in the phase space occurs for a
critical surface in the parameter space. Consequently, the energy spectrum
in the bosonic sector becomes infinitely degenerate, while the degeneracy
in the fermionic sector gets enhanced by a factor of two for each pair
of reduced canonical coordinates. The disparity in the enhancement of
degeneracy between the bosonic and the fermionic sector can be attributed
to the fact that bosonic quantum number can take any non-negative integer
values, while a fermionic quantum number can take only two values.

The plan of the paper is the following. We introduce the noncommutative
algebra among different phase-space operators in the next section. We
also construct ${\cal{N}}=2$ supersymmetric quantum mechanics on the
noncommutative space for an arbitrary superpotential. The super-Hamiltonian
for the particular case of superoscillator in $N$ dimensions is obtained.
In Sec. III, we study the two dimensional superoscillator in detail and obtain
its energy spectrum. Different limiting cases
including the `inverted superoscillator' is discussed in this section.
The superoscillator in three dimensions is discussed in Sec. IV. Finally,
we conclude in Sec. V. In Appendix A, the procedure for diagonalizing the
three dimensional bosonic Hamiltonian is discussed. In Appendix B, the matrix
representation of the fermionic Hamiltonian for $N=3$ is given.

\section{General formulation}
Consider the noncommutative algebra,
\be
\left [ \hat{x}_i , \hat{x}_j \right ] = i \theta_{ij}, \ \ \
\left [ \hat{p}_i , \hat{p}_j \right ] = i B_{ij}, \ \ \
\left [ \hat{x}_i , \hat{p}_j \right ] = i \delta_{ij} + i \left ( 1
- \delta_{ij} \right ) \ C_{ij}, \ \
i, j=1, 2, \dots, N,
\label{eq0}
\ee
\noindent where $\theta_{ij}$ and $B_{ij}$ are real, antisymmetric matrices 
and are independent of the hermitian operators $\hat{x}_i, \hat{p}_i$.
The diagonal elements of the matrix $C_{ij}$ are taken to be zero and its
off-diagonal elements do not depend on the space and momentum coordinates.
We introduce $2 N$ elements $\xi_i$ satisfying the following
real Clifford algebra,
\be
\{\xi_i, \xi_j \} = 2 g_{ij}, \ \ g_{ij}=  \pmatrix{ {-{\cal{I}}}& {0}\cr
{0} & {{\cal{I}}}\cr},
\label{cliff}
\ee
\noindent where ${\cal{I}}$ is an $N \times N$ identity matrix. The signature
of the metric $g_{ij}$ is such that the square of the $\xi_i$'s is equal to
$1$ or $-1$ depending on whether  $ i > N$ or $ i \leq N$, respectively. In
particular,
\be
\xi_{N+i}^2 = - \xi_{i}^2=1, \ \ i=1,2, \dots N.
\ee
\noindent In this paper, we use a particular matrix representation of the
Clifford algebra so that the following additional relations are also satisfied,
\be
\xi_i^{\dagger} = - \xi_i, \ \ \xi_{N+i}^{\dagger} = \xi_{N+i},
\label{chk}
\ee
\noindent where $X^{\dagger}$ denotes the hermitian adjoint of $X$. Such a
matrix representation of the Clifford algebra is required to construct
hermitian Hamiltonian within our approach.
We further introduce the hermitian operator $\gamma_5$,
\be
\gamma_5 =  \xi_1 \xi_2 \dots \xi_{2 N-1} \xi_{2 N},
\ee
\noindent which anticommutes with all the $\xi_i$'s and $\gamma_5^2=1$.
The elements of the Clifford algebra $\xi_i, \xi_{N+i}$ commute with the
noncommutative bosonic variables $\hat{x}_i$ and $\hat{p}_i$,
\be
\left [ \hat{x}_i, \xi_j \right ] = \left [ \hat{p}_i, \xi_j \right ] =0, \ \
\left [ \hat{x}_i, \xi_{N+j} \right ] = \left [ \hat{p}_i, \xi_{N+j} \right ]
=0, \ \ \forall \ i, j.
\ee
\noindent It naturally follows that $\gamma_5$ also commutes with all
the bosonic coordinates.

We now introduce the supercharges $Q_1$ and $Q_2$\cite{susy,dhowker},
\be
Q_1 =  \frac{1}{\sqrt{2}} \sum_{i=1}^N \left [ - i \ \xi_i \ \hat{p}_i +
\xi_{N+i} \ \hat{W}_i \right ],\ \
Q_2 = - i \gamma_5 Q_1.
\label{q1s1}
\ee
\noindent The superpotential $\hat{W}_i$ are real functions of the
noncommutative coordinate $\hat{x}_i$ and in general $\left [ \hat{W}_i,
\hat{W}_j \right ] \neq 0$. Note that both $Q_1$ and $Q_2$ are constructed
to be hermitian operators for real $\hat{W}_i$. The supercharges $Q_1$ and
$Q_2$ satisfy the following standard superalgebra,
\be
\{Q_{\alpha}, Q_{\beta} \} = 2 \delta_{\alpha, \beta} H, \ \
\left [H, Q_{\alpha} \right ] = 0, \ \ \alpha, \beta = 1, 2,
\label{algebra}
\ee
\noindent where the Hamiltonian $H$ is given by,
\be
H = \frac{1}{2} \sum_{i=1}^N  \left ( \hat{p}_i^2 + \hat{W}_i^2 \right )
-\frac{i}{4} \sum_{i,j=1}^N \left ( B_{ij} \xi_i \xi_j +
2 \xi_i \xi_{N+j} \left [ \hat{p}_i, \hat{W}_j \right ] + 
i \ \xi_{N+i} \xi_{N+j} \left [ \hat{W}_i, \hat{W}_j \right ] \right ).
\label{hami}
\ee
\noindent The Hamiltonian $H$ is hermitian, since it is given by the square
of the hermitian operator $Q_1(Q_2)$. The hermiticity of $H$ can also be
checked explicitly using the equation (\ref{chk}). The term containing
 $B_{ij}$ arises due to the noncommutativity
among momentum operators. Similarly, the last term in $H$ that is proportional
to $ [ \hat{W}_i, \hat{W}_j ]$ arises due to the noncommutativity among space
coordinates. Such a term is absent for supersymmetric quantum mechanics on
commutative space.

A few comments are in order at this point.\\
(i) If we allow $\theta_{ij}, B_{ij}$ and $C_{ij}$ to be functions of the
noncommutative coordinates $\hat{x}_i, \hat{p}_i$ instead of c-number matrices,
the whole analysis up to the construction of the super-Hamiltonian (\ref{hami})
remains valid. It is worth mentioning here that the Jacobi identities severely
restrict the choice of the operators $\theta_{ij}, B_{ij}, C_{ij}$ for such
more general theories.\\
(ii) In the standard construction of supersymmetric quantum mechanics on
the commutative space, one usually introduces fermionic variables $\psi_i$
and its conjugate $\psi_i^{\dagger}$,
\be
\psi_i = \frac{i}{2} \left (\xi_i - \xi_{N+i} \right ), \ \
\psi_i^{\dagger} = \frac{i}{2} \left (\xi_i + \xi_{N+i} \right ),
\ee
\noindent so that the eigenstates can be labeled in terms of the total
fermion number $N_F=\sum_i \psi_i^{\dagger} \psi_i$. However, one can
check that $H$ contains term of the form $\psi_i \psi_j, \psi_i^{\dagger}
\psi_j^{\dagger}$ for $B_{ij}, \theta_{ij} \neq 0$, implying $N_f$ is not
a conserved quantity.\\
(iii) Let us define another set of supercharges $q_1$ and $q_2$,
\be
q_1 = - \frac{1}{\sqrt{2}} \sum_{i=1}^N \left ( \xi_{N+i} \hat{p}_i
+ i \xi_i \hat{W}_i \right ), \ \ q_2= - i \gamma_5 q_1.
\ee
\noindent These two supercharges satisfy the standard superalgebra
(\ref{algebra}) with the Hamiltonian $h \equiv q_1^2=q_2^2$,
\be
h = \frac{1}{2} \sum_{i=1}^N  \left ( \hat{p}_i^2 + \hat{W}_i^2 \right )
+ \frac{i}{4} \sum_{i,j=1}^N \left ( B_{ij} \xi_{N+i} \xi_{N+j} -
2 \xi_i \xi_{N+j} \left [ \hat{p}_j, \hat{W}_i \right ] +
i \ \xi_{i} \xi_{j} \left [ \hat{W}_i, \hat{W}_j \right ] \right ).
\label{kami}
\ee
\noindent In the commutative limit $B_{ij}=\theta_{ij}=C_{ij}=0$, the two
Hamiltonian $H$ and $h$ are identical. Moreover, in the same limit,
the pair of charges $(Q_1,q_1)$ and the pair $(Q_2, q_2)$ satisfy
the superalgebra (\ref{algebra}) separately. Although, $(Q_1, Q_2, q_1, q_2)$
do not close under an enlarged ${\cal{N}}=4$ superalgebra,
we have the freedom of choosing any pair of the supercharges $\{(Q_1, Q_2),
(q_1, q_2), (Q_1, q_1), (Q_2, q_2) \}$ for a given ${\cal{N}}=2$
super-Hamiltonian $H=h$ on the commutative space. However, if we take any of
the parameters $\theta_{ij}, B_{ij}, C_{ij}$ non-zero, such a freedom is
completely lost. We are compelled to choose either the pair $(Q_1, Q_2)$ or
$(q_1, q_2)$ and of-course, in general, the super-Hamiltonian $H$ and $h$ are
not identical.

\subsection{Superoscillators}

We will be working with $Q_1, Q_2$ and $H$ in the rest of the paper. We also
consider $\theta_{ij}, B_{ij}, C_{ij}$ as constant matrices from now onward.
For the case of superoscillator, we choose,
\be
\hat{W}_i = \omega \hat{x}_i.
\ee
\noindent The Hamiltonian now reads,
\bea
H & = & H_b - H_f\nonumber \\
H_b & \equiv & \frac{1}{2} \sum_{i=1}^N  \left ( \hat{p}_i^2 +
\omega^2 \hat{x}_i^2 \right ),\nonumber \\
H_f & \equiv & \frac{\omega}{2} \left ( \sum_{i=1}^N \xi_i \xi_{N+i}
- \sum_{i \neq j} \xi_{N+i} \xi_j C_{ij} \right )
+ \frac{i}{4} \sum_{i,j=1}^N \left ( B_{ij} \xi_i \xi_j
- \omega^2 \theta_{ij} \xi_{N+i} \xi_{N+j} \right ).
\label{dami}
\eea
\noindent 
In the limit of $B_{ij}, \theta_{ij}, C_{ij} \rightarrow 0$, the Hamiltonian
of the superoscillator on commutative space is recovered.
Note that $H_b$ is a function of the noncommutative co-ordinates
and the momenta only, whereas $H_f$ is solely expressed in terms of the
elements of the Clifford algebra. This implies that $H_b$ and $H_f$ can
be diagonalized separately.

A comment is in order at this point. The $N$ dimensional superoscillator
is described in terms of $2 N$ elements of the Clifford algebra (\ref{cliff}).
So, the Hamiltonian $H_f$ can be expressed in terms of the linear combination
of the $ N (2 N-1)$ generators $\Sigma_{ij}^{1,2,3}$,
\be
\Sigma_{ij}^1 = \frac{i}{4} \left [ \xi_i, \xi_j \right ], \ \
\Sigma_{ij}^2 = \frac{i}{4} \left [ \xi_{N+i}, \xi_{N+j} \right ], \ \
\Sigma_{ij}^3 = \frac{i}{4} \left [ \xi_i, \xi_{N+j} \right ],
\ee
\noindent of the group $SO(N,N)$ of rank $N$. Thus, in general, the eigenvalues
of $H_f$ can be expressed in terms of $N$ quantum numbers. Further, in the
matrix representation of the Clifford algebra (\ref{cliff}), both
the generators $\Sigma_{ij}^{1,2,3}$ and the Hamiltonian $H_f$ can be expressed
in terms of $2^N \times 2^N$ dimensional matrices. Thus, each of the $N$
quantum numbers can take only two values.

\section{Two dimensional superoscillator}

We now specialize in this section to the noncommutative plane for which the
antisymmetric matrices $B_{ij}$ and $\theta_{ij}$ can be parametrized
in terms of single parameters $B$ and $\theta$, respectively. In particular,
\be
C_{12} \equiv \phi_1, \ \ C_{21} \equiv -\phi_2, \ \
B_{ij} \equiv \epsilon_{ij} B, \ \ \theta_{ij} \equiv \epsilon_{ij} \theta,
\ \ i, j= 1, 2.
\ee
\noindent With this choice for $B_{ij}$ and $\theta_{ij}$, there are many
physically equivalent representations\cite{poly,3d3} of the algebra
(\ref{eq0}) in terms of commutative canonically conjugate variables $x_i$ and
$p_i$ satisfying
\be
[x_i, x_j]=0, [p_i, p_j]=0, [x_i, p_j]=i \delta_{ij}.
\ee
\noindent As shown in Ref. \cite{poly,3d1,3d3}, the Hamiltonian $H_b$ with $N=2$
can be equivalently written as a two-dimensional anisotropic oscillator.
In particular,
\bea
H_b & = & \frac{1}{2} \left [ \Omega_+ \left ( p_1^2 + x_1^2 \right )
+ \Omega_- \left ( p_2^2 + x_2^2 \right ) \right ],\nonumber \\
2 \Omega_{\pm} & = & \sqrt{\left (\omega^2 \theta -B \right )^2 +
4 \omega^2 + \omega^2 \left ( \phi_1 + \phi_2 \right)^2 } \pm
\sqrt{\left (\omega^2 \theta + B \right )^2 +
\omega^2 \left ( \phi_1 - \phi_2 \right )^2}, \ \ \kappa > 0,\nonumber \\ 
2 \Omega_{\pm} & = &\pm \sqrt{\left (\omega^2 \theta -
B \right )^2 + 4 \omega^2 + \omega^2 \left ( \phi_1 + \phi_2 \right)^2 } +
\sqrt{\left (\omega^2 \theta + B \right )^2 +
\omega^2 \left ( \phi_1 - \phi_2 \right )^2}, \kappa < 0,\nonumber \\
\Omega_+ & = & \sqrt{\left (\omega^2 \theta + B \right )^2 +
\omega^2 \left (\phi_1 -\phi_2 \right )^2}, \ \ \Omega_-=0, \kappa=0, 
\eea
\noindent where the positive and the negative values of the parameter
$\kappa \equiv \omega^2 ( 1 - B \theta + \phi_1 \phi_2) $ correspond to two
different phases of the noncommutative oscillators, the critical value being
$\kappa=0$. The energy eigenvalues $E_b$ of $H_b$ is,
\be
E_b = \left ( n_+^b + \frac{1}{2} \right ) \Omega_+
+ \left ( n_-^b + \frac{1}{2} \right ) \Omega_-, 
\label{evb}
\ee
\noindent where the quantum numbers $n_{\pm}^b$ can take any non-negative
integer values.

In order to diagonalize $H_f$, we use the following matrix representation
of the Clifford algebra,
\be
\xi_1 = i \sigma_1 \otimes \sigma_2, \ \
\xi_2 = i \sigma_2 \otimes \sigma_2, \ \
\xi_3 = - \sigma_3 \otimes \sigma_2, \ \
\xi_4 = I \otimes \sigma_3,
\ee
\noindent where $\sigma_{1,2,3}$ are the three Pauli matrices and $I$ is a
$2 \times 2$ identity matrix. The Hamiltonian $H_f$ is a $4 \times 4$ matrix,
\be
H_f = \left ( \matrix{  {\alpha} & {\beta}\cr \\ {\beta} & {\alpha}} \right ), \ \
\alpha \equiv \left ( \matrix{  {\frac{B}{2}} & {\frac{\omega}{2} ( i + \phi_1)}\cr \\
{\frac{\omega}{2} ( -i + \phi_1)} & {-\frac{B}{2}}} \right ), \ \
\beta \equiv \left ( \matrix{  {-\frac{\omega^2 \theta}{2}} & {\frac{\omega}{2} ( i +
\phi_2)}\cr \\ {\frac{\omega}{2} ( -i + \phi_2)}  & {\frac{\omega^2 \theta}{2}}} \right ).
\ee
\noindent The eigenvalues of $H_f$ are,
\be
\{ -\frac{1}{2}(\Omega_+ - \Omega_-),
\frac{1}{2}(\Omega_+ - \Omega_-), -\frac{1}{2}(\Omega_+ + \Omega_-),
\frac{1}{2}(\Omega_+ + \Omega_-) \}.
\ee
\noindent For $\Omega_-=0$, there are only two independent eigenvalues
$\pm \frac{1}{2} \Omega_+$, each of them being doubly degenerate. 
The frequency $\Omega_-$ vanishes only in the critical phase $\kappa=0$.
Separate discussions are needed for the critical and the non-critical
phases.

\subsection{Critical phase $\kappa=0$}
The frequency $\Omega_-=0$ in the critical phase.
The eigenvalues $E_f$ of $H_f$ can be written in a closed form as,
\be
-E_f = \frac{1}{2} n_+^f \  n_-^f \Omega_+, \ \ n_{\pm}^f=-1, 1.
\label{evf}
\ee
\noindent The nonlinear dependence on the `fermionic' quantum numbers 
$n_{\pm}^f$ is due to the doubly degenerate eigenvalues
$\pm \frac{1}{2} \Omega_+$. Combining Eqs. (\ref{evb}) and (\ref{evf}),
we find the energy eigenvalues $E$ of the super-Hamiltonian $H$,
\bea
E & = & E_b - E_f\nonumber \\
& = & \left ( n_+^b + \frac{1}{2} n_+^f n_-^f + \frac{1}{2} \right ) \Omega_+,
\label{ener}
\eea
\noindent where the quantum numbers $n_{\pm}^b$ can take any non-negative
integer values, while the `fermionic' quantum numbers $n_{\pm}^f$ can take
only two values, either $-1$ or $1$.

For $\Omega_- =0$, a reduction in the phase-space occurs. The phase-space
variables $p_2$ and $x_2$ completely decouple from $H_b$.
The energy eigenvalues $E_b$ of the bosonic sector and consequently $E$
becomes infinitely degenerate. Similarly, the four eigenvalues of
$H_f$ reduces to two with each of them being doubly degenerate.
The groundstate energy $E=0$ is obtained, either for
(i) $n_+^b=0, n_+^f=1, n_-^f=-1$ or (ii)$n_+^b=0, n_+^f=-1, n_-^f=1$ .
The supersymmetry is unbroken.

The condition $\kappa=0$ has two solutions, (a) $\omega=0$ and
(b) $1-B \theta + \phi_1 \phi_2=0$. In the limit $\omega \rightarrow 0$,
$ \Omega_+=B$ and the eigenvalue equation of $H$ reduces to that of Pauli
equation on the two dimensional
commutative space. Note that the Pauli Hamiltonian in our case has higher
symmetry than the one usually considered in the literature \cite{km},
since we have taken $4 \times 4$ dimensional matrix representation of the
Clifford algebra instead of the familiar $2 \times 2$ dimensional
representation. So, the degeneracy structure of the energy spectrum is
different from Ref. \cite{km}.

\subsection{ Non-critical phase $\kappa \neq 0$}

In the non-critical phase, both $\Omega_{\pm} \neq 0$. The eigenvalue $E_f$
\be
-E_f = \frac{1}{2} \left ( n_-^f \Omega_- + n_+^f \Omega_+ \right ),
\ee
\noindent is identical to that of an anisotropic `fermionic' oscillator with
two degrees of freedom. The eigenvalue $E$ of the superHamiltonian $H$ is,
\be
E= \left ( n_+^b + \frac{1}{2} n_+^f + \frac{1}{2} \right ) \Omega_+ 
+ \left ( n_-^b + \frac{1}{2} n_-^f + \frac{1}{2} \right ) \Omega_-.
\label{alt}
\ee
\noindent The spectrum $E$ in Eq. (\ref{alt}) is identical to that of an
anisotropic superoscillator on the commutative plane. The groundstate energy
$E=0$ is obtained for $n_{\pm}^b=0, n_{\pm}^f=-1$. All excited states
are paired together. The supersymmetry is unbroken.

The isotropic superoscillator can
be obtained as a special case, when the parameters $B, \theta, \omega,
\phi_{1,2}$ satisfy the following two relations simultaneously,
\be
\omega^2 \theta + B =0, \ \ \phi_1=\phi_2, \ \ \kappa > 0.
\label{rela}
\ee
\noindent For $\kappa < 0$, there is no `isotropic limit'.
The commutative limit, $B, \theta, \phi_{1,2} \rightarrow 0$
is of-course a solution of the above equation (\ref{rela}) for which
$\Omega_{\pm}=\omega^2$.
Note that the isotropic superoscillator can be obtained as a special
case only when both $\theta$ and $B$ are either zero or nonzero
simultaneously.

\subsection{Inverted superoscillator on the noncommutative plane}
In the commutative space, the inverted harmonic oscillator
\be
{\cal{H}} = \frac{1}{2} \sum_{i=1}^2 \left ( p_i^2 -
\bar{\omega}^2 x_i^2 \right )
\ee
\noindent with $\bar{\omega}$ real, is unbounded from below and there is no
well defined groundstate. However, for the noncommutative case with both $B$
and $\theta$ being nonzero, $\bar{H}_b$ admits well defined
groundstate\cite{invert}. This result can be understood intuitively as follows.
The fundamental uncertainty relations are modified due to the
non-commutativity and lower bounds appearing in these relations are expected to
explicitly depend on $\theta$ and $B$. These parameters act as a kind of
regulators/cut-offs to the singular potential for some ranges of their allowed
values. Consequently, the potential effectively becomes bounded from below and
admits well defined groundstate.

To analyze a similar situation in case of the supersymmetric inverted
oscillator, we make the transformation $\omega \rightarrow i \bar{\omega}$ in
all previous calculations in this section and identify any operator/function
$A$ undergoing such transformation as $\bar{A}$. The transformation
$\omega \rightarrow i \bar{\omega}$ amounts to taking a complex superpotential.
The supercharges $\bar{Q}_1, \bar{Q}_2$ and the Hamiltonian $\bar{H}_f$ are no
more hermitian operators.
Further, unlike the standard supersymmetric theory, the supersymmetric
Hamiltonian $\bar{H}$ is not a semi-positive definite quantity. These features
are common to inverted superoscillator both on the commutative as well as
on the non-commutative space. The inverted superoscillator on the commutative
space do not have a well defined ground state. However, in spite of the
undesirable features like non-hermiticity and unboundedness from below, we will
show that the spectrum of the inverted superoscillator on the noncommutative
space is indeed real. It is known that a certain class of non-hermitian
Hamiltonian on the ordinary commutative space admits real
spectra\cite{complex,me}. To the best of our knowledge, the discussion in this
section is the first example of a non-hermitian Hamiltonian admitting real
spectra in the context of a supersymmetric theory as well as on a
noncommutative space.

We find the frequencies of $\bar{H}_b$, 
\bea
2 \bar{\Omega}_{\pm} & = & \sqrt{(\bar{\omega}^2 \theta - B)^2 -
\bar{\omega}^2 (\phi_1 - \phi_2)^2 + 4 \bar{\kappa} } \pm
\sqrt{ \left (\bar{\omega}^2 \theta - B \right )^2
-\bar{\omega}^2 \left ( \phi_1 - \phi_2 \right )^2 }, \ \bar{\kappa} > 0,
\nonumber \\
2 \bar{\Omega}_{\pm} & = & \pm \sqrt{(\bar{\omega}^2 \theta - B)^2 
- \bar{\omega}^2 (\phi_1 - \phi_2)^2 + 4 \bar{\kappa} } +
\sqrt{ \left (\bar{\omega}^2 \theta - B \right )^2
-\bar{\omega}^2 \left ( \phi_1 - \phi_2 \right )^2 }, \ \bar{\kappa} < 0,
\nonumber \\
\bar{\Omega}_+ & \equiv & \Omega = \sqrt{(\bar{\omega}^2 \theta - B)^2 -
\bar{\omega}^2 (\phi_1 - \phi_2)^2 }, \ \ \Omega_-=0, \ \ \bar{\kappa}=0, \eea
\noindent where $\bar{\kappa}=-\bar{\omega}^2 ( 1 - B \theta +
\phi_1 \phi_2 )$. We discuss the three cases separately.\\
(i) $\bar{\kappa} < 0$ : The commutative limit,
$( \phi_{1,2}, \theta, B) \rightarrow 0$, can be taken only in the phase
$\bar{\kappa} < 0$. We find a complex $\bar{\Omega}_{\pm}$ in this limit,
implying that the solutions are not stable. The commutative limit with nonzero
magnetic field, i.e. $( \phi_{1,2}, \theta) \rightarrow 0 \ \& \ B \neq 0$, also
belongs to this phase. The frequencies $\bar{\Omega}_{\pm}$ are complex
for $B^2 < 4 {\mid \bar{\kappa} \mid}$ and $\Omega_-$ is negative for
$B^2 > 4 {\mid \bar{\kappa} \mid}$ signifying physically non-acceptable
non-normalizable solutions. In-fact, in the phase $\bar{\kappa} < 0$, there
are no physically acceptable solutions, since $\bar{\Omega}_{\pm}$ are either
complex or negative.\\
(ii) $\bar{\kappa} > 0$: There are regions in the parameter space for which
physically acceptable solutions with well defined ground state energy can be
found. In particular, for $ ( \bar{\omega}^2 \theta - B)^2 \geq
\bar{\omega}^2 (\phi_1 - \phi_2)^2 $, both $\bar{\Omega}_{\pm}$ are real
and positive. When the bound is saturated, we get an isotropic oscillator with
$\bar{\Omega}_+=\bar{\Omega}_-=\sqrt{\bar{\kappa}}$. The energy spectrum is
given by
$\bar{E}=( n_-^b + \frac{1}{2} n_-^f + \frac{1}{2} ) \bar{\Omega}_-
+ ( n_+^b + \frac{1}{2} n_-^f + \frac{1}{2} ) \bar{\Omega}_+ $ with the zero
groundstate energy.\\ 
(iii) $ \bar{\kappa}=0$: At the critical point $\bar{\kappa}=0$ there is a
reduction in the phase space. For $ ( \bar{\omega}^2 \theta - B)^2 >
\bar{\omega}^2 (\phi_1 - \phi_2)^2 $, $\Omega$ is real and positive.
The energy spectrum is,
$\bar{E}= ( n_+^b + \frac{1}{2} n_+^f + \frac{1}{2} ) {\Omega}$
with a well defined ground state $\bar{E}_0=0$. The energy spectrum is
infinitely degenerate.


\section{Three dimensional superoscillator}

In this section, we discuss three dimensional superoscillator on the
noncommutative space. For simplicity, we choose
$C_{ij} = 0 \ \forall \ i \ \& \ j$ so that the canonical commutation
relations between the non-commutative coordinates and the momenta now read,
$[\hat{x}_i, \hat{p}_j ] = i \delta_{ij}$. Each of the antisymmetric matrices 
$B_{ij}$ and $\theta_{ij}$ is parameterized in terms of three 
parameters. In particular,
\be
B_{ij} \equiv \epsilon_{ijk} B_k, \ \
\theta_{ij} \equiv \epsilon_{ijk} \theta_k.
\label{para}
\ee
\noindent We also define,
\be
B \equiv \sqrt{B_1^2 + B_2^2 + B_3^2}, \ \
\theta \equiv \sqrt{\theta_1^2 + \theta_2^2 + \theta_3^2},
\ee
\noindent for convenience.
The bosonic part of the Hamiltonian is still given by $H_b$ in
(\ref{dami}) with $N=3$. As is well known\cite{3d1,3d3} and described in
Appendix A, the Hamiltonian $H_b$ can be expressed solely in terms of
canonically conjugate variables $x_i$ and $p_i$ as an anisotropic oscillator
in three dimensional commutative space. In particular,
\be
H_b= \frac{1}{2} \left [ \Omega_+ \left ( p_1^2 + x_1^2 \right )
+ \Omega_- \left ( p_2^2 + x_2^2 \right )
+ \Omega_0 \left ( p_3^2 + x_3^2 \right ) \right ],
\label{boson}
\ee
\noindent where $\pm \Omega_{\pm}, \pm \Omega_0$ are the six eigenvalues of
the $6 \times 6$ matrix $ i M$ given in appendix A with
$C_{ij} = 0 \ \forall \ i \ \& \ j$. There are two phases of the theory
characterized by $\kappa > 0$ or $ \kappa < 0$ with $\kappa=0$ being the
critical phase, where $\kappa \equiv \Omega_+ \Omega_- \Omega_0$. The
frequencies $\Omega_{\pm}, \Omega_0$ are always positive for the range of
parameters that is determined depending on the particular phase
($\kappa=0, \kappa > 0, \kappa < 0$) in which the superoscillators
are being considered.

The Hamiltonian $H_f$ with the parametrization (\ref{para}) now reads,
\be
H_f = \frac{\omega}{2} \sum_{i=1}^3 \xi_i \xi_{3+i}
+ \frac{i}{4} \sum_{i,j=1}^3 \epsilon_{ijk} \left ( B_k \xi_i \xi_j
- \omega^2 \theta_k \xi_{3+i} \xi_{3+j} \right ).
\ee
\noindent
We choose the matrix representation of the Clifford algebra (\ref{cliff}),
\bea
&& \xi_1 = i \sigma_3 \otimes \sigma_2 \otimes I, \ \ \
\xi_2 = i\sigma_1 \otimes \sigma_2 \otimes I, \ \ \
\xi_3 = i \sigma_2 \otimes I \otimes \sigma_3,\nonumber \\
&&
\xi_4 = \sigma_2 \otimes I \otimes \sigma_1, \ \ \ \
\xi_5 = I \otimes \sigma_3 \otimes \sigma_2, \ \ \ \
\xi_6 = I \otimes \sigma_1 \otimes \sigma_2.
\eea
\noindent The Hamiltonian $H_f$ in terms of these matrix representation
of the elements of the Clifford algebra is given in Appendix B.

A general analysis of the spectrum of $H$ involves the diagonalization of the
$6 \times 6$ matrix $M$ and $8 \times 8$ matrix $H_f$. The eigenvalues of both
of these matrices come into positive-negative pairs. Thus, all the eigenvalues
of $H$ can be obtained analytically by solving a cubic and a quartic
characteristic equation, respectively. However, the final expressions for
the frequencies $\Omega_i$ are quite messy and complicated. We present here
results for a few specific choices of parameters with $\omega > 0$. The results
for $\omega < 0$ can be obtained trivially by considering the negative
pairs of the frequencies $\Omega_{\pm}, \Omega_0$. The results for the
purely bosonic model has been obtained previously for some of the cases
considered below,
while the results for the fermionic part and hence, the complete supersymmetric
model are obtained for the first time in this paper. We would like to mention
here that the supersymmetry is unbroken for all the cases considered below.\\
(a) ${\bf B_i = 0, \  \forall \ i}$: The noncommutativity among the momentum
operators are switched off. There is only one phase characterized by
$\kappa= \omega^3 > 0$ for positive $\omega$. The frequencies are,
\be
\Omega_0= \omega, \ \
\Omega_{\pm}= \omega \sqrt{t \pm \sqrt{t^2-1}}, \ \
t \equiv 1+\frac{1}{2} \omega^2 \theta^2.
\ee
\noindent There is no limit for nonzero $\theta_i$'s for which $\Omega_+
=\Omega_-$. The eight eigenvalues of $H_f$,
\be
\{ \frac{\omega}{2} \left ( 1 \pm \sqrt{2 + 2 t} \right), \ \
- \frac{\omega}{2} \left ( 1 \pm \sqrt{2 + 2 t} \right), \ \
\frac{\omega}{2} \left ( 1 \pm \sqrt{2 t-1} \right), \ \
- \frac{\omega}{2} \left ( 1 \pm \sqrt{2 t-1} \right) \}
\ee
\noindent can be written in a closed form through the introduction of
three quantum numbers $n_{\pm}^f, n_0^f$,
\be
- E_f = \frac{1}{2} \left ( n_-^f \Omega_- + n_+^f \Omega_+ + n_0^f \Omega_0
\right ), \ \ n_{\pm}^f, n_0^f= -1, 1.
\label{close}
\ee
\noindent Thus, the eigenspectrum of $H$,
\be
E=\left ( n_+^b + \frac{1}{2} n_+^f + \frac{1}{2} \right ) \Omega_+
+\left ( n_-^b + \frac{1}{2} n_-^f + \frac{1}{2} \right ) \Omega_-
+ \left ( n_0^b + \frac{1}{2} n_0^f + \frac{1}{2} \right ) \Omega_0,
\label{sun}
\ee
\noindent is that of an anisotropic superoscillator on the three
dimensional commutative space. The bosonic quantum numbers $n_{\pm}^b,
n_0^b$ can take any non-negative integer values. The supersymmetric
groundstate $E=0$ is obtained for $n_{\pm}^b=n_0^b=0, n_{\pm}^f=n_0^f=-1$.\\
(b) ${\bf B_i  = - \omega^2 \theta_i, \ \forall \ i}$:
For this choice, two of the three frequencies become equal,
\be
\Omega_0 = \omega, \ \ 
\Omega_+ = \Omega_- \equiv \Omega = \omega \sqrt{1 +
\omega^2 \theta^2},
\label{iso}
\ee
\noindent producing an isotropic oscillator on the x-y plane plus an
oscillator along the z-direction. There is again only one phase $\kappa > 0$
for positive $\omega$. The eight eigenvalues of $H_f$ can
be expressed in a closed form as in (\ref{close}) with the frequencies
given by Eq. (\ref{iso}). The eigenspectrum of $H$,
\be
E=\left ( n_+^b + n_-^b + \frac{1}{2} n_+^f + \frac{1}{2} n_-^f +
1 \right ) \Omega +\left ( n_0^b + \frac{1}{2} n_0^f
+ \frac{1}{2} \right ) \Omega_0,
\ee
\noindent has higher level of degeneracy compared to (\ref{sun}). The
supersymmetric groundstate is obtained for all bosonic quantum numbers
being zero and all the fermionic quantum numbers equal to $-1$.\\
(c) ${\bf B_i  = \omega^2 \theta_i, \ \forall \ i}$:
There are two phases in the theory characterized by $\kappa =
\omega^3 (1 - \omega^2 \theta^2)$. The frequencies in these two phases are,
\bea
&& \Omega_0 = \omega, \ \
\Omega_{\pm} = \omega \left ( 1 \pm \omega \theta \right ),\ \
\kappa > 0,\nonumber \\ 
&& \Omega_0 = \omega, \ \
\Omega_{\pm} = \omega \left (\pm 1 + \omega \theta \right ), \ \
\kappa < 0.
\label{2p}
\eea
\noindent There is no isotropic point $ \Omega_- = \Omega_+ $ for
$\theta_i \neq 0$. The eigenspectrum of $H$ is given by Eq. (\ref{sun})
with the frequencies are as in Eq. (\ref{2p}).\\
(d) {\bf Critical point $\kappa=0$}:
The critical point $\kappa=0$ separating the two phases of the
superoscillators is determined from the relation,
$Det(M)=\omega^6 ( 1 - \sum_{i=1}^3 \theta_i B_i)=0$. There are two cases,
(i)$\sum_{i=1}^3 \theta_i B_i=1$ and (ii) $\omega=0$.\\
(i)$\sum_{i=1}^3 \theta_i B_i=1$: Without any loss of generality, 
we choose a particular solution, $B_i = \theta_i \theta^{-2}$. 
The phase-space is reduced to $4$ dimensions,
since one of the frequency vanishes. In particular,
\be
\Omega_0=0, \ \ 
\Omega_+=\omega, \ \
\Omega_-=\theta^{-1} \left ( 1 + \omega^2 \theta^2 \right ). 
\ee
\noindent 
The bosonic spectrum becomes infinitely degenerate.
At the critical point, there are four independent eigenvalues of
$H_f$ with each of them being doubly degenerate,
\be
\{ \pm \frac{\omega}{2} + \frac{1}{2 \theta} \left ( 1 +
\theta^2 \omega^2 \right ),
\pm \frac{\omega}{2} - \frac{1}{2 \theta} \left ( 1 +
\theta^2 \omega^2 \right ) \}.
\ee
\noindent The eigenvalues of $H$ can be written in a closed form as,
\be
E = \left ( n_+^b + \frac{1}{2} n_0^f n_+^f + \frac{1}{2} \right ) \Omega_+
+ \left ( n_-^b + \frac{1}{2} n_0^f n_-^f + \frac{1}{2} \right ) \Omega_-.
\ee
\noindent The $E=0$ groundstate is obtained for (i) $n_+^b=0, n_0^f=1,
n_{\pm}^f=-1$ and (ii) $n_+^b=0, n_0^f=-1, n_{\pm}^f=1$. The whole spectrum
including the groundstate is of-course infinitely degenerate due to the
reduction in the phase-space. 

\noindent (ii) $\omega=0$: The eigenvalue equation of the Hamiltonian $H$
is that of a three dimensional Pauli equation. There are further reductions
in the
phase space, compared to the case described above. Four out of the six
phase space variables decouple from the dynamics. In particular,
\be
\Omega_{\pm}=0, \ \ \Omega_0= B.
\ee
\noindent The eigenvalues of $H_f$ are $\pm \frac{1}{2} \Omega_0$ with
each of them having four-fold degeneracy. The eigenspectrum of $H$
can be written as, 
\be
E=(n_0^b + \frac{1}{2} n_0^f n_-^f n_+^f + \frac{1}{2}) \Omega_0.
\ee
\noindent The $E=0$ groundstate can be obtained for (i) $n_0^b=0, n_0^f=-1,
n_{\pm}^f=1$, (ii) $n_0^b=0, n_0^f=-1, n_{\pm}^f=-1$, (iii) $n_0^b=0, n_0^f=1,
n_{\pm}^f=\pm 1$ and (iv) $n_0^b=0, n_0^f=1, n_{\pm}^f= \mp 1 $.
The whole spectrum including the groundstate is of-course infinitely
degenerate due to the reduction in the phase-space. 

\section{Summary \& Discussions}

We have constructed supersymmetric quantum mechanics on the noncommutative
space in terms of ${\cal{N}}=2$ real supercharges. This construction is
valid in any arbitrary dimensions for arbitrary superpotential. Further,
the same construction is valid for a very large class of noncommutativities
among phase-space operators. The non-commutativity among the space coordinates
and/or among momentum coordinates restricts the number of independent ways
one can construct ${\cal{N}}=2$ supersymmetric theory for a given
superpotential.

We have studied the noncommutative superoscillators in two and three dimensions
in some detail and obtained their eigenspectra analytically. The spectrum of
the noncommutative superoscillator in
two(three) dimensions is identical to an anisotropic oscillator in the
commutative two(three) space dimensions. There is a critical surface
in the parameter space for which there is reduction in the phase-space.
Consequently, the spectrum due to the bosonic sector become infinitely
degenerate, while the degeneracy in the fermionic sector is doubled. This
is because the bosonic quantum numbers can take any non-negative integer
values, while the fermionic quantum numbers can take only two values.

We have studied the `inverted superoscillator' on the noncommutative
plane. The corresponding Hamiltonian is neither hermitian nor a semi-positive
definite operator. Inspite of these, we have found the spectrum to be real, 
positive and with a well defined ground state for a few specific choices of
the parameters. For these choices of parameters,
there is no smooth limit to the ordinary commutative space. Thus, the
energy spectrum with a well defined ground state for an
`inverted superoscillator' is purely the effect of noncommutativity.

\acknowledgments{ It is a great pleasure to acknowledge several useful
discussions with Prof. Avinash Khare. I also thank him for a very critical
reading of the manuscript. This work is supported (DO No. SR/FTP/PS-06/2001)
by SERC, DST, Govt. of India through the Fast Track Scheme for Young
Scientists:2001-2002.}

\appendix{
\section{Mapping of $H_b$ to an anisotropic oscillator on
commutative space for $N=3$}

In this Appendix, we describe a general procedure for mapping the
non-commutative Hamiltonian $H_b$ to an equivalent Hamiltonian of anisotropic
oscillators on the commutative space. The discussion is based on Refs.
\cite{3d3,3d1,book}.

Define a $6$ dimensional vector $U=(\omega \hat{x}_1, \hat{p}_1,
\omega \hat{x}_2, \hat{p}_2, \omega \hat{x}_3, \hat{p}_3)$ in the phase
space. The algebra in (\ref{eq0}) can be written as,
\be
[U_I, U_J] = i M_{IJ}, \ \ I, J=1, 2, \dots 6,
\ee
\noindent where the $6 \times 6$ dimensional matrix $M$ is given by,
\be
M=\left (\matrix{{0} & {\omega} & {\omega^2 \theta_3} & {\omega C_{12}} &
{-\omega^2 \theta_2} & {\omega C_{13}}\cr \\
{-\omega} & {0} & {-\omega C_{21}} & {B_3} & {-\omega C_{31}} & {-B_2}\cr \\
{-\omega^2 \theta_3} & {\omega C_{21}} & {0} & {\omega} &
 {\omega^2 \theta_1} & {\omega C_{23}}\cr \\
{-\omega C_{12}} & {-B_3} & {-\omega} &{0} & {-\omega C_{32}} & {B_1}\cr \\
{\omega^2 \theta_2} & {\omega C_{31}} & {-\omega^2 \theta_1} & {\omega C_{32}}
& {0} & {\omega}\cr \\
{-\omega C_{13}} & {B_2} & {-\omega C_{23}} & {-B_1} & {-\omega} &
{0}} \right ).
\ee
\noindent There exists an orthogonal transformation such that the matrix $M$
can be block-diagonalized as\cite{book},
\be
R^T M R = \left ( \matrix{ {D_1} & 0 &\cr \\
{0} & {D_2} & {0}\cr \\ {0} & {0} & {D_3}} \right ), \ \ \
D_i \equiv \left ( \matrix{ {0} & {\Omega_i}\cr \\
{-\Omega_i} & {0}} \right ),
\ee
\noindent using a $O(6)$ rotational matrix $R$ and its transpose $R^T$,
where $\pm {\Omega_i}$ are the six eigenvalues of $i M$. The matrix $R$
is unique modulo $O(6)$ rotations. The transformed variables in the
phase space,
$(x_1, p_1, x_2, p_2, x_3, p_3) \equiv (\frac{u_1}{\sqrt{{\mid \Omega_1 \mid}}},
\frac{u_2}{\sqrt{{\mid \Omega_1 \mid}}} 
\frac{u_3}{\sqrt{{\mid \Omega_2 \mid}}} 
\frac{u_4}{\sqrt{{\mid \Omega_2 \mid}}}
\frac{u_5}{\sqrt{{\mid \Omega_3 \mid}}} 
\frac{u_6}{\sqrt{{\mid \Omega_3 \mid}}})$
with $u = R^T U$ satisfy the usual canonical commutation relations
\be
[x_i, x_j]=0, \ \ [p_i, p_j]=0, \ \ [x_i, p_j]=i \delta_{ij}.
\ee
\noindent The Hamiltonian $H_b=\frac{1}{2} \sum_{I=1}^6 U_I^2=
\frac{1}{2} \sum_{I=1}^6 u_I^2$, because of the $O(6)$ invariance.
We get back the Hamiltonian (\ref{boson}) once we express the variables
$u_I$ in terms of $x_i$ and $p_i$.

For $N=2$, We can truncate the matrix $M$ to a $4 \times 4$ matrix by
taking the first four row and columns only, i.e. $M_{IJ}$ with
$I,J=1,2,3,4$. Identifying $\theta_3= \theta$, $B_3=B$, $C_{12}=\phi_1$
and $c_{21}=-\phi_2$, all the relevant results of Sec. III can be obtained.

\section{$H_f$ in matrix representation for $N=3$}

The Hamiltonian $H_f$ in terms of the matrix representation of the Clifford
algebra has the following form,
\be
H_f=\left ( \matrix{ {\Gamma_1} & {\eta}\cr \\ {\eta^{\dagger}} & {\Gamma_2}}
\right ),
\ee
\noindent where the $4 \times 4$ matrices $\Gamma_1$, $\Gamma_2$ and $\eta$
are,
\bea
&& \Gamma_1 \equiv \left ( \matrix{ {0} & {-\gamma_3^+} & {-\gamma_1^+} &
{-\gamma_2^-}\cr \nonumber \\
{\gamma_3^+} & {0} & {-\gamma_2^+} & {\gamma_1^-}\cr \nonumber \\
{\gamma_1^+} & {\gamma_2^+} & 0 & {-\gamma_3^-}\cr \nonumber \\
{\gamma_2^-} & {-\gamma_1^-} & {\gamma_3^-} & {0}} \right ),\nonumber \ \
\Gamma_2 \equiv \left ( \matrix{ {0} & {-\gamma_3^-} &{\gamma_1^-} &
{\gamma_2^-}\cr \nonumber \\
{\gamma_3^-} & {0} & {\gamma_2^+} & {-\gamma_1^{+}}\cr \nonumber \\
{-\gamma_1^-} & {- \gamma_2^+} & {0} & {-\gamma_3^+}\cr \nonumber \\
{-\gamma_2^{-}} & {\gamma_1^+} & {\gamma_3^+} & {0}} \right ),\cr \nonumber \\
\ \ \nonumber \\
&& \eta\equiv \left ( \matrix{{0} & {0} & {0} & {-\frac{i}{2} \omega}\cr
\nonumber\\
{0} & {0} & {\frac{i}{2} \omega} & {0}\cr \nonumber \\
{0} & {\frac{i}{2} \omega} & {0} & {0}\cr \nonumber \\
{\frac{3 i}{2} \omega} & {0} & {0} & {0}} \right ), \hspace{.5in} 
\gamma_i^{\pm} \equiv \frac{i}{2} ( B_i \pm
\theta_i \omega^2 ).\cr \nonumber \\
\cr \nonumber \ \
\eea
\noindent Note that,
\be
\left ( \gamma_i^{\pm} \right )^* = - \gamma_i^{\pm},
\ee
\noindent where $(X)^*$ denotes the complex conjugate of $X$.
The hermiticity of $H_f$ can now be checked easily.
}


\begin{thebibliography}{99}
\bibitem{connes} A. Connes, Noncommutative zeometry, Academic Press,
London, (1994).

\bibitem{sw} N. Seiberg and E. Witten, JHEP {\bf 9909} (1999) 032
[hep-th/9908142].

\bibitem{pert} S. Minwalla, M. V. Raamsdonk, N. Seiberg,
JHEP {\bf 0002} (2000) 020 [hep-th/9912072].

\bibitem{npert} R. Gopakumar, S. Minwalla and A. Strominger,
JHEP {\bf 0005} (2000) 020 [hep-th/0003160].

\bibitem{rev1} M. R. Douglas and N. A. Nekrasov, Rev. Mod. Phys. {\bf 73}
(2001) 977 [hep-th/0106048].

\bibitem{rev2} J. A. Harvey, Komaba Lectures on Noncommutative Solitons and
D-Branes, hep-th/0102076.

\bibitem{rev3} R. J. Szabo, Int. J. Mod. Phys. {\bf A19} (2004) 1837
[physics/0401142]; Phys. Rep. {\bf 378} (2003) 207
[hep-th/0109162].

\bibitem{bra}  V. O. Rivelles, hep-th/0305122. 

\bibitem{tama} D. Klemm, S. Penati and L. Tamassia, Class. Quant. Grav.
{\bf 20} (2003) 2905 [hep-th/0104190].

\bibitem{seib} N. Seiberg, JHEP {\bf 0306} (2003) 010 [hep-th/0305248].

\bibitem{fr} C. Duval and P. Horvathy, Phys. Lett. {\bf B479} (2000) 284
[hep-th/0002233]; J. Phys. {\bf A34} (2001) 10097 [hep-th/0106089];
P. Horvathy, Ann. Phys. {\bf 299} (2002) 128 [hep-th/0201007].

\bibitem{lamb} M. Chaichian, M. M. Sheikh-Jabbari and A. Tureanu,
Phys. Rev. Lett. {\bf 86} (2001) 2716 [hep-th/0010175].


\bibitem{masud} M. Chaichian, P. Presnajder, M. M. Sheikh-jabbari and A.
Tureanu, Phys. Lett. {\bf B527} (2002) [hep-th/0012175].

\bibitem{poly} V. P. Nair and A. P. Polychronakos, Phys. Lett. {\bf B505}
(2001) 267 [hep-th/0011172].

\bibitem{3d1} A. Hatzinikitas and I. Smyrnakis, J. Math. Phys. {\bf 43}
(2002) 113 [hep-th/0103074].

\bibitem{others} A. Jellal, J. Phys. {\bf A34} (2001) 10159 [hep-th/0105303];
S. Bellucci, A.  Nersessian and C. Sochichiu, Phys. Lett. {\bf B52} (2001)
345 [hep-th/0106138];
A. Smailagic and E.  Spalluci, Phys. Rev. {\bf D65} (2002) 107701
[hep-th/0108216];
B. Muthukumar and P. Mitra, Phys. Rev. {\bf D66} (2002) 027701 [hep-th/0204149].

\bibitem{rabi} R. Banerjee, Mod. Phys. Lett. {\bf A17} (2002) 631 [0106280].

\bibitem{3d2} A. Smailagic and E. Spallucci, J. Phys. {\bf A365} (2002)
L363 [hep-th/0205242].

\bibitem{3d3} L. Jonke and S. Meljanac, Eur. Phys. J. {\bf C29} (2003) 433
[hep-th/0210042]; I. Dadic, L. Jonke and S. Meljanac, hep-th/0301066.

\bibitem{invert} S. Bellucci, Phys. Rev {\bf D67} (2003) 105014
[hep-th/0301227].

\bibitem{cks} F. Cooper, A. Khare and U. Sukhatme, Phys. Rep. {\bf 251}
(1995) 267.

\bibitem{star} T. Curtright, D. Fairlie and C. K. Zachos,
Phys. Rev. {\bf D58} (1998) 025002 [hep-th/9711183].

\bibitem{khare} E. Harikumar, V. Sunil Kumar and A. Khare, Phys. Lett.
{\bf B589} (2004) 155 [hep-th/0402064].

\bibitem{pp} P. Pouliot, Class. Quant. Grav. {\bf 21} (2004)
145 [hep-th/0306261].


\bibitem{susy} P. K. Ghosh, Nucl. Phys. {\bf B681} (2004) 359 [hep-th/0309183].

\bibitem{dhowker} E. D'Hoker and L. Vinet, Comm. Math. Phys. {\bf 97}
(1985) 391.

\bibitem{km} M. de Crombrugghe and V. Rittenberg, Ann. Phys. {\bf151} (1983)
99; A. Khare and J. Maharana, Nucl. Phys. {\bf B244} (1984) 409.

\bibitem{complex} C. M. Bender, D. C. Brody and H. F. Jones, Am. J. Phys.
{\bf 71} (2003) 1095 [hep-th/0303005]; C.M. Bender and S. Boettcher, Phys. Rev.
Lett. {\bf 80} (1998) 5243.

\bibitem{me} P. K. Ghosh, quant-ph/0501087.

\bibitem{book} D. McDuff and D. Salamon, Introduction to Symplectic Topology,
Oxford Science Publications (1998).
\end{thebibliography}
\end{document}